\documentclass[nofootinbib,aps,amssymb,floatfix,superscriptaddress,preprintnumbers,twocolumn]{revtex4}
\usepackage{dcolumn}
\usepackage{graphicx}
\usepackage{bm}
\usepackage{amsmath}
\usepackage{amssymb}
\usepackage{amsfonts}
\usepackage{float}
\usepackage{hyperref}
\usepackage{dsfont}  
\usepackage{slashed}  
\usepackage{booktabs}
\usepackage{multirow}
\usepackage{subfigure}
\usepackage[sort&compress]{natbib}
\usepackage{xcolor}
\usepackage{ulem}

\newcommand{\be}{\begin{equation}}  
\newcommand{\ee}{\end{equation}}  
\newcommand{\beq}{\begin{eqnarray}} 
\newcommand{\eeq}{\end{eqnarray}}

\newcommand{\bea}{\begin{eqnarray}}
\newcommand{\eea}{\end{eqnarray}}

\begin{document}

\title{Revealing chiral-odd two-meson generalized distribution amplitudes\\ in  $e^- e^+ \to (\pi \pi) (\pi \pi)$ reactions}
\author{Shohini Bhattacharya}
\email{shohinib@uconn.edu}
\affiliation{Department of Physics, University of Connecticut, Storrs, CT 06269, USA}

\author{Renaud Boussarie}
\email{renaud.boussarie@polytechnique.edu}

\author{Bernard Pire}
\email{bernard.pire@polytechnique.edu}
\affiliation{CPHT, CNRS, \'Ecole polytechnique, Institut Polytechnique de Paris, 91128 Palaiseau, France}

\author{Lech Szymanowski}
\email{lech.szymanowski@ncbj.gov.pl}
\affiliation{National Centre for Nuclear Research (NCBJ), 02-093 Warsaw, Poland}

\begin{abstract}
We demonstrate that chiral-odd dimeson generalized distribution amplitudes (CO-GDAs)—nonperturbative objects encoding the transition of a quark–antiquark pair into two mesons— can
be accessed in high-energy $e^- e^+$ annihilation into two meson pairs, each with a relatively low invariant mass. While chiral-even GDAs contribute to the leading one-photon amplitude, the chiral-odd sector enters via two-photon exchange. We show that the interference between these amplitudes leads to  
specific effects which may be measurable at BES III or future tau-charm factories. This work opens a direct path to experimentally probing the long-missing chiral-odd sector of meson structure—specifically, the spin-orbit correlation  in a  spin-zero meson, in some contexts referred to as anomalous tensorial magnetic moment.
\end{abstract}
\maketitle

{\bf 1. Introduction.}
The quest to image the internal structure of hadrons has led to the development of powerful nonperturbative tools such as generalized parton distributions (GPDs) \cite{Diehl:2003ny, Belitsky:2005qn}, which encode correlations between longitudinal momentum and transverse position of partons. Their crossed-channel counterparts, generalized distribution amplitudes (GDAs), offer an equally rich but comparatively less explored avenue: they describe the hadronization of a quark-antiquark or gluon pair into an exclusive meson pair and are accessible in processes such as $\gamma^* \gamma \to \pi \pi$~\cite{Diehl:1998dk,Polyakov:1998ze}. Much like GPDs, GDAs provide tomographic information—now for mesons—and carry unique sensitivity to hadron dynamics such as partial-wave structure, resonance content, and spin-dependent correlations~\cite{Pire:2002ut,Diehl:2000uv}.

Chiral-even dipion GDAs have already been probed experimentally in $e^+e^-$ collisions at Belle~\cite{Belle:2015oin}, enabling the first data-driven analyses~\cite{Kumano:2017lhr} and demonstrating their viability as phenomenological objects. However, the chiral-odd sector of GDAs remains entirely unexplored. These distributions, which are the crossing analogues of transversity GPDs~\cite{Diehl:2001pm}, encode tensor structures and transverse spin correlations that are inaccessible through chiral-even observables. Their measurement would not only complete the GDA framework by incorporating the long-missing chiral-odd sector, but also reveal how transverse spin degrees of freedom are encoded in exclusive hadronization processes.

From a phenomenological standpoint, chiral-odd GDAs are expected to contribute to exclusive dipion production with transversely polarized photons or appear in observables sensitive to interference with chiral-even amplitudes, as we demonstrate explicitly in this work. Establishing theoretical control over these distributions and identifying observables where they can be cleanly accessed are essential steps toward a full three-dimensional understanding of hadron structure.

In this work, we demonstrate, for the first time, that chiral-odd dimeson GDAs may be accessed in high-energy $e^+e^-$ annihilation into two meson pairs, each with a relatively low invariant mass. Chiral-even GDAs dominate the process through single-photon exchange, whereas chiral-odd contributions arise via two-photon exchange. We demonstrate that their interference gives rise to transverse-momentum-dependent observables, providing experimental access to the chiral-odd sector through measurable effects at BES III and future high luminosity $e^+e^-$ colliders such as the super tau-charm factory (STCF)~\cite{Achasov:2023gey, Ai:2025xop}. We estimate the size of these effects. Our study thus paves the way for a potential experimental probe of the tensor structure and the spin–orbit correlation~\cite{Lorce:2025ayr} in a spin-zero meson, sometimes referred to~\cite{Hagler:2009ni} as the anomalous tensorial magnetic moment of the pion in analogy with the case of the nucleon~\cite{Brodsky:2002pr}.

{\bf 2. Chiral-even and chiral-odd GDAs.} As in the chiral-even case, the leading-twist chiral-odd $\pi\pi$ GDAs decompose into isoscalar and isovector components. For a charged pion pair, only the isovector GDA contributes and is defined as follows:
\begin{align}
 & \langle\left\{ \begin{array}{c}
\pi^{+}(p_{1})\\
\pi^{-}(p_{1})
\end{array}\right\} \pi^{0}(p_{2})|\left\{ \begin{array}{c}
\bar{u}(v)\\
\bar{d}(v)
\end{array}\right\} [v,0]\gamma^{\lambda}\left\{ \begin{array}{c}
d(0)\\
u(0)
\end{array}\right\} |0\rangle\nonumber \\
 & =\frac{p_{1}^{\lambda}+p_{2}^{\lambda}}{\sqrt{2}}\int_{0}^{1}\!{\rm d}z\,{\rm e}^{iz(p_{1}+p_{2})\cdot v}\Phi_{{\rm ce}}^{V}(z,\zeta,s),\label{eq:pi+pi0GDA}
\end{align}
\begin{align}
 & \langle\left\{ \begin{array}{c}
\pi^{+}(p_{1})\\
\pi^{-}(p_{1})
\end{array}\right\} \pi^{0}(p_{2})|\left\{ \begin{array}{c}
\bar{u}(v)\\
\bar{d}(v)
\end{array}\right\} [v,0]i\sigma^{\mu\nu}\left\{ \begin{array}{c}
d(0)\\
u(0)
\end{array}\right\} |0\rangle\nonumber \\
 & =\frac{p_{1}^{\mu}p_{2}^{\nu}-p_{1}^{\nu}p_{2}^{\mu}}{m_{\pi}}\int_{0}^{1}\!{\rm d}z\,{\rm e}^{iz(p_{1}+p_{2})\cdot v}\Phi_{{\rm co}}^{V}(z,\zeta,s).\label{eq:PhiVcoDef}
\end{align}
Here, $z$ denotes the light-cone momentum fraction of the quark, $\zeta$ is the light-cone momentum fraction of the final-state charged mesons and $s=(p_1+p_2)^2$ is the usual Mandelstam variable. We note that the isovector chiral-even GDAs (chiral-odd GDAs), $\Phi_{{\rm ce}}^{V}$ ($\Phi_{{\rm co}}^{V}$) are identical for the $\pi^+ \pi^0$ and $\pi^- \pi^0$ channels, as shown in Eqs.~(\ref{eq:pi+pi0GDA}) and~(\ref{eq:PhiVcoDef}).

The QCD evolution of GDAs is governed~\cite{Diehl:1998dk,Polyakov:1998ze} by the Efremov–Radyushkin–Brodsky–Lepage (ERBL) equations~\cite{Efremov:1979qk,Lepage:1979zb}. Chiral-odd quark GDAs evolve independently of gluon GDAs, just as transversity GPDs evolve independently of gluon distributions.
The solution to the ERBL evolution equations is typically expressed in terms of Gegenbauer polynomials \( C_n^k(2z - 1) \) and Legendre polynomials \( C_l^{1/2}\left(\frac{2\zeta - 1}{\sqrt{1 - 4m_\pi^2/s}}\right) \)~\cite{Polyakov:1998ze}. In the isovector sector, the asymptotic form reads:
\begin{eqnarray}
  \Phi_{\text{asy}}^{V}(z, \zeta, s, \mu) &=& 6\, B_{01}^0(s, \mu)\, z(1 - z) \frac{2\zeta - 1}{\sqrt{1 - 4m_\pi^2/s}}\,,
\end{eqnarray}
with the scale dependence encoded in
\begin{equation}
B_{nl}^0(s, \mu) = B_{nl}^0(s, \mu_0) \left[ \frac{\alpha_s(\mu^2)}{\alpha_s(\mu_0^2)} \right]^{(\gamma_n - \gamma_0)/(2\beta_0)}\,.
\end{equation}
The asymptotic \((z, \zeta)\)-dependence of chiral-odd GDAs is the same as that of chiral-even GDAs, but the anomalous dimensions differ: \((\gamma_0, \gamma_1) = (0,\, 25/9)\) for the chiral-even case, and \((\gamma_0, \gamma_1) = (8/3,\, 104/9)\) for the chiral-odd case.

The coefficient $B_{01}(s,\mu)$ is closely related to the timelike form factor of the pion, as the first moment of the isovector GDA reads~\cite{Polyakov:1998ze}:
\begin{equation}
    \int dz \Phi^V(z,\zeta,s) =(2\zeta-1)F^\pi(s)\,.
\end{equation}
Retaining only the leading (asymptotic) term in the evolution-driven expansion of the GDA yields the equality $B_{01}(s,\mu) = F^\pi(s)$.
The normalization factors \( B_{nl}(s, \mu) \) are generally complex, and distinguishing their phases is often essential. In particular, they read
\begin{eqnarray}
 B_{01} (s,\mu)&=& \left|B_{01} (s,\mu)\right| e^{i\delta_V(s)}~~,\\ ~~ B_{T01} (s,\mu) &=& \left|B_{T01} (s,\mu) \right|e^{i\delta_V^T(s)}\,.
\end{eqnarray}
Contrary to its chiral-even analog, the normalization factor \( B_{T01}(s,\mu) \) is unknown and related to the tensor form factor of the pion. Since the phases are understood as arising from final-state \(\pi\pi\) interactions—driven by the phases accompanying the Breit-Wigner parametrizations of \(s\)-channel resonances such as the \(\rho\) and \(\rho'\)—one may, as a first step, assume that the phases of the chiral-odd and the corresponding chiral-even GDAs are equal:
\begin{equation}
    \delta_V^T = \delta_V\,.
    \label{phases}
\end{equation}
This assumption, however, must ultimately be tested against experimental data. Given that the \(s\)-dependence of the normalization factors \( B_{nl}(s,\mu) \) can be derived from the phase \(\delta_V\) via an Omnes representation~\cite{Omnes:1958hv}, Eq.~\eqref{phases} implies
\begin{equation}
    B_{T01}(s,\mu) = K\, B_{01}(s,\mu)\,,
    \label{norm}
\end{equation}
where \(K\) is an unknown proportionality constant. Determining the magnitude of this factor \(K\), as well as accessing the relative phase between the chiral-odd and chiral-even GDAs, constitutes a central objective of the experimental strategy proposed here. We will make use of Eq.~(\ref{norm}) in the numerical analysis below, using input for $K$ motivated by lattice results.
As chiral-odd GDAs also contribute to multibody \(B\)-meson decays~\cite{Cheng:2019hpq,Yan:2025ocu}, the results may have significant implications for the study of nonperturbative QCD dynamics in heavy-flavor physics.

{\bf 3. Kinematics for the process $\boldsymbol{e^{-}e^{+}\to (\pi^+\pi^0)(\pi^-\pi^0)}$.}
\begin{figure}
 \hspace{-0.5cm} \includegraphics[width=0.4\textwidth]{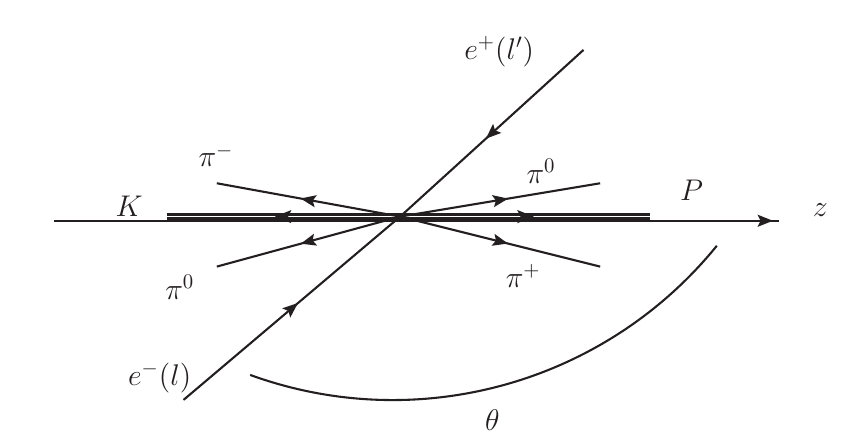}
 
 \vspace{-1.1cm} \hspace{-0.5cm}
\includegraphics[width=0.5\textwidth]{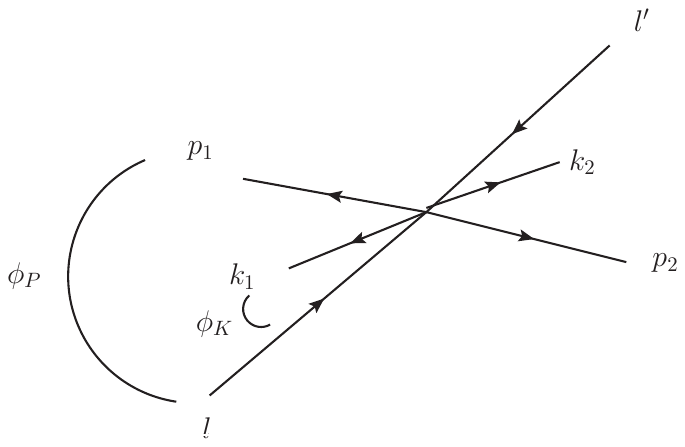}
\vspace{-8cm}
\caption{Kinematics of the $e^{-}(\ell)e^{+}(\ell')\to (\pi^+(p_1)\pi^0(p_2))(\pi^-(k_1)\pi^0(k_2))$ process  (upper panel) and its transverse view (lower panel) defining the azimuthal angles $\phi_P, \phi_K$.}
\label{proces-view}
\end{figure}
We want to study the process $e^{-}(\ell)e^{+}(\ell^{\prime})\rightarrow (\pi^{+}(p_{1})\pi^{0}(p_{2})) (\pi^{-}(k_{1})\pi^{0}(k_{2}))$
in the kinematics where the two pion pairs have small center-of-mass energies $s_{P}=(p_{1}+p_{2})^{2}$
and $s_{K}=(k_{1}+k_{2})^{2}$: $s_{P},s_{K}\ll s =(\ell +\ell')^2$. We work in the
overall center of mass frame.  In exact kinematics, 
we define two
lightlike vectors $P$, 
$K$ such that 
\begin{equation}
(P+K)^{2}\equiv s_{0}=\frac{s-s_{P}-s_{K}+\sqrt{\lambda(s,s_{P},s_{K})}}{2}\label{eq:}
\end{equation}
with the Källén function $\lambda(s_{1},s_{2},s_{3})\equiv s_{1}^{2}+s_{2}^{2}+s_{3}^{2}-2s_{1}s_{2}-2s_{1}s_{3}-2s_{2}s_{3}$.
The transverse direction is defined with respect to these two vectors, and transverse components will be denoted by a $\perp$ subscript in Minkowski space, or in boldface in Euclidean space.
We parameterize the momenta as follows,\footnote{Here and throughout this article, we denote \(\bar{\alpha} \equiv 1 - \alpha\) for any scalar \(\alpha\).} with the skewness variables defined as \(\zeta_P = \frac{p_1 \cdot K}{P \cdot K}\) and \(\zeta_K = \frac{k_1 \cdot P}{P \cdot K}\):
\begin{equation}
\left\{ \begin{array}{c} p_1 \\ p_2 \end{array}  \right\}= \left\{ \begin{array}{c} \zeta_P  \\ \bar\zeta_P \end{array}  \right\} P + \left\{ \begin{array}{c} \bar\zeta_P  \\ \zeta_P \end{array}  \right\} \frac{s_{P}}{s_{0}}K  + \left\{ \begin{array}{c} 1  \\ -1 \end{array}  \right\}  \frac{\delta_{P\perp}}{2}
\end{equation}
for one pion pair, and
\begin{equation}
\left\{ \begin{array}{c} k_1 \\ k_2 \end{array}  \right\}= \left\{ \begin{array}{c} \zeta_K  \\ \bar\zeta_K \end{array}  \right\} K + \left\{ \begin{array}{c} \bar\zeta_K  \\ \zeta_K \end{array}  \right\} \frac{s_{K}}{s_{0}}P  + \left\{ \begin{array}{c} 1  \\ -1 \end{array}  \right\} 
\frac{\delta_{K\perp}}{2}\,,
\end{equation}
for the other. The lepton momenta are parameterized as 
\begin{equation}
\left\{ \begin{array}{c} l \\ l' \end{array}  \right\}= \left\{ \begin{array}{c} u  \\ \bar u \end{array}  \right\}(1+\frac{s_K}{s_0}) P + \left\{ \begin{array}{c} \bar u \\ u \end{array}  \right\} (1+\frac{s_{P}}{s_{0}})K  + \left\{ \begin{array}{c} 1  \\ -1 \end{array}  \right\} 
\ell_{\perp}\,.
\end{equation}
Up to lepton mass corrections, transverse vectors are normalized as follows:
\begin{equation}
\boldsymbol{\ell}^{2}=u\bar{u}s,\,\boldsymbol{\delta}_{P}^{2}=4\zeta_{P}\bar{\zeta}_{P}s_{P}-4m_{\pi}^{2},\,\boldsymbol{\delta}_{K}^{2}=4\zeta_{K}\bar{\zeta}_{K}s_{K}-4m_{\pi}^{2}\,,\label{eq:TransVecSquares}
\end{equation}
and $u$ is related to the angle $\theta$ between the electron beam and the dihadron $p_1+p_2$ momenta (see Fig.~\ref{proces-view}) by the relation 
\begin{equation}
 u=(1-\cos \theta)/2\,.   
\end{equation}
Finally, we define a two-dimensional transverse vector \(\boldsymbol{n}\) such that \(\left(\frac{\boldsymbol{\ell}}{|\boldsymbol{\ell}|}, \frac{\boldsymbol{n}}{|\boldsymbol{\ell}|}\right)\) forms a right-handed orthonormal basis of the transverse plane. We then define the angles \(\phi_P\) and \(\phi_K\), as shown in Fig.~\ref{proces-view}, such that
\begin{align}
\boldsymbol{\delta}_{P} & =\frac{\left|\boldsymbol{\delta}_{P}\right|}{\left|\boldsymbol{\ell}\right|}(\cos\phi_{P}\boldsymbol{\ell}+\sin\phi_{P}\boldsymbol{n})\label{eq:PhiPdef}\,,\\
\boldsymbol{\delta}_{K} & =\frac{\left|\boldsymbol{\delta}_{K}\right|}{\left|\boldsymbol{\ell}\right|}(\cos\phi_{K}\boldsymbol{\ell}+\sin\phi_{K}\boldsymbol{n})\label{eq:PhiKdef}.
\end{align}

{\bf 4. The factorized scattering amplitude for $\boldsymbol{e^-e^+ \to (\pi^+\pi^0) (\pi^-\pi^0)}$.}
\begin{figure*}[htbp!]
\centering
\includegraphics[width=10cm]{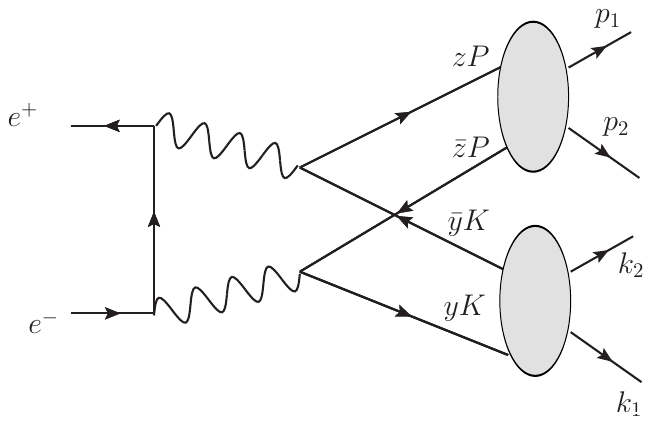}
\hspace{-3cm}
\includegraphics[width=10cm]{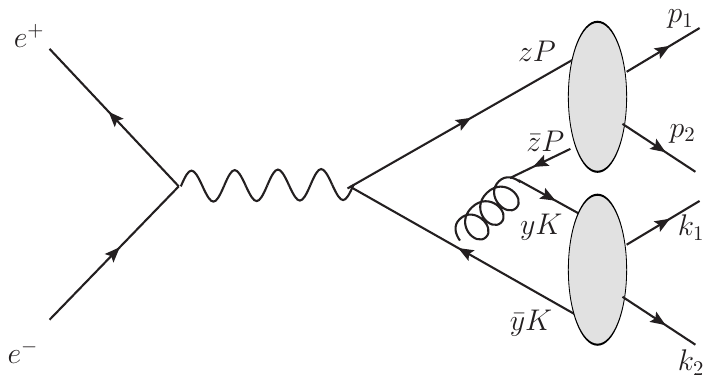}
\vspace{-9cm}
\caption{Feynman diagrams for leading-order $(\pi^+ \pi^0)(\pi^- \pi^0)$ production. Left: two-photon exchange (the diagram with interchanged photon lines is implied). Right: one-photon exchange (three additional diagrams are to be added).}
\label{fig:C-evendiagrams}
\end{figure*}
The amplitude for the production of $\pi^{+}(p_{1})\pi^{0}(p_{2})$ and $\pi^{-}(k_{1})\pi^{0}(k_{2})$ pairs receives contributions from both single-photon (Charge-odd) exchange and double-photon (Charge-even) exchange mechanisms; see Fig.~\ref{fig:C-evendiagrams}.  For the single-photon exchange case, factorization has been formally proven~\cite{Efremov:1979qk,Lepage:1979zb} for the production of two hadrons. For the production of hadron pairs, the factorization theorem (which is by definition independent from the nature of the hadronic final states) applies in our kinematics where the invariant mass of the pairs are small. For the double-photon exchange contribution, we justify factorization by noting that both GDAs and meson DAs are confined to the ERBL region, so the hard coefficient function remains unaffected by their replacement. Only the end-point regions require careful treatment, as in standard meson production processes. Given the large-angle \(2 \to 2\) kinematics and the point-like nature of all other vertices, our process thus falls within the ERBL factorization framework~\cite{Efremov:1979qk,Lepage:1979zb}. 

From now on, we use the convention ${\cal S}=1+i(2\pi)^{4}\delta^{4}(p_{1}+p_{2}+k_{1}+k_{2}-\ell-\ell^{\prime}){\cal T}$,  $Q_{u}=2/3,Q_{d}=-1/3$ are the electric charge fractions of the quarks; $e^2 =4\pi \alpha_{em}$ and $g^2 = 4\pi \alpha_S$.
The single photon contribution contains only chiral-even distributions and reads
:
\begin{equation}
{\cal T}_{1}=i\frac{Q_{u}-Q_{d}}{u-\bar{u}}\frac{e^{2}g^{2}}{8N_{c}s^{2}}(\bar{v}_{\ell^{\prime}}\slashed{\ell}_{\perp}u_{\ell})\Psi_{1}^{V}(\zeta_{P},s_{P},\zeta_{K},s_{K})\label{eq:T1resgen},
\end{equation}
while the two photon exchange amplitude gets a chiral-even contribution (we remove the arguments of the $\Psi$ convolutions for the sake of readability):
\begin{equation}
    {\cal T}_{2{\rm ce}} =\frac{e^{4}Q_{u}Q_{d}}{8N_{c}s^{2}}(\bar{v}_{\ell^{\prime}}\slashed{\ell}_{\perp} \label{eq:T2ceresgen}u_{\ell})\Psi_{2{\rm ce}}^{V}
\end{equation}
and a chiral-odd contribution:
\begin{equation}
{\cal T}_{2{\rm co}}=\frac{e^{4}Q_{u}Q_{d}}{2N_{c}s^{2}}\bar{v}_{\ell^{\prime}}(\ell_{\perp}^{\alpha}\gamma_{\perp}^{\beta}+\ell_{\perp}^{\beta}\gamma_{\perp}^{\alpha}-g_{\perp}^{\alpha\beta}\slashed{\ell}_{\perp})u_{\ell}\frac{\delta_{P\perp}^{\alpha}\delta_{K\perp}^{\beta}}{4m_{\pi}^{2}}\Psi_{2{\rm co}}^{V}\label{eq:T2coresgen}
\end{equation}
with the GDAs contributing through their convolutions with the coefficient functions defined as~\footnote{In our amplitudes, two GDAs enter: one for the $(p_1,p_2)$ pion pair and one for the $(k_1,k_2)$ pair. We will denote the associated momentum fractions respectively as $y$ and $z$.}:
\begin{equation}
\Psi_{1}^{V}=\int_{0}^{1}\frac{{\rm d}y}{y\bar{y}}\Phi_{{\rm ce}}^{V}(y,\zeta_{P},s_{P})\int_{0}^{1}\frac{{\rm d}z}{z\bar{z}}\Phi_{{\rm ce}}^{V}(z,\zeta_{K},s_{K})\,,\label{eq:Psi1def}
\end{equation}
\begin{align}
&\Psi_{2{\rm ce}}^{V}(\zeta_{P},s_{P},\zeta_{K},s_{K})  =\\&\int_{0}^{1}{\rm d}y\int_{0}^{1}{\rm d}z\frac{(y\bar{y}+z\bar{z})\Phi_{{\rm ce}}^{V}(y,\zeta_{P},s_{P})\Phi_{{\rm ce}}^{V}(z,\zeta_{K},s_{K})}{y\bar{y}z\bar{z}(y\bar{z}-yu-\bar{u}\bar{z}+i0)(z\bar{y}-u\bar{y}-\bar{u}z+i0)}\,,\nonumber\label{eq:Psi2cedef}
\end{align}
\begin{align}
&\Psi_{2{\rm co}}^{V}(\zeta_{P},s_{P},\zeta_{K},s_{K})  =\\&\int_{0}^{1}{\rm d}y\int_{0}^{1}{\rm d}z\frac{(-1+\bar{y}z+y\bar{z})\Phi_{{\rm co}}^{V}(y,\zeta_{P},s_{P})\Phi_{{\rm co}}^{V}(z,\zeta_{K},s_{K})}{y\bar{y}z\bar{z}(y\bar{z}-yu-\bar{u}\bar{z}+i0)(z\bar{y}-u\bar{y}-\bar{u}z+i0)}\, . \label{eq:Psi2codef}
\end{align}
These one-photon and two-photon amplitudes interfere in the calculations of various observables that we now address.

{\bf 5. The  cross section for $\boldsymbol{e^-e^+ \to (\pi^{+}\pi^{0}) (\pi^{-}\pi^{0})}$ and numerical results.} We now focus on the $(\pi^{+}(p_{1})\pi^{0}(p_{2})) (\pi^{-}(k_{1})\pi^{0}(k_{2}))$ channel, which proves to be the most promising isospin configuration for accessing the chiral-odd GDAs
The differential cross-section reads:
\begin{align}
&\frac{{\rm d}\sigma^{+0;-0}}{{\rm d}s_{P}{\rm d}s_{K}{\rm d}u{\rm d}\zeta_{P}{\rm d}\phi_{P}{\rm d}\zeta_{K}{\rm d}\phi_{K}}  = \nonumber\\ 
& \frac{u\bar{u}\alpha_{{\rm em}}^{2}}{2^{9}\pi^{3}N_{c}^{2}s^{3}} \left|\frac{\alpha_{s}(Q_{u}-Q_{d})}{4}\Psi_{1}^{V}-\alpha_{{\rm em}}(u-\bar{u})Q_{u}Q_{d}\Psi_{2{\rm ce}}^{V}\right|^{2}\nonumber\\
 & +\frac{u\bar{u}\alpha_{{\rm em}}^{3}Q_{u}Q_{d}}{2^{10}\pi^{3}N_{c}^{2}s^{3}}(u-\bar{u})\cos(\phi_{P}+\phi_{K})\frac{\left|\boldsymbol{\delta}_{P}\right|\left|\boldsymbol{\delta}_{K}\right|}{4m_{\pi}^{2}}\times\nonumber \\
 & {\rm Re}\left\{ \left[\frac{\alpha_{s}(Q_{u}-Q_{d})}{4}\Psi_{1}^{V}-\alpha_{{\rm em}}(u-\bar{u})Q_{u}Q_{d}\Psi_{2{\rm ce}}^{V}\right]\Psi_{2{\rm co}}^{V\ast}\right\} \nonumber \\
 & +\frac{u\bar{u}\alpha_{{\rm em}}^{4}Q_{u}^{2}Q_{d}^{2}}{2^{13}\pi^{3}N_{c}^{2}s^{3}}[1-4u\bar{u}\cos^{2}(\phi_{P}+\phi_{K})]\frac{\boldsymbol{\delta}_{P}^{2}}{4m_{\pi}^{2}}\frac{\boldsymbol{\delta}_{K}^{2}}{4m_{\pi}^{2}}\left|\Psi_{2{\rm co}}^{V}\right|^{2} .
  \label{eq:dSigma+0-0}
\end{align}
In Eq.~\ref{eq:dSigma+0-0}, the first two terms arise from chiral-even GDAs and their interference with chiral-odd contributions, while the third term---quadratic in the chiral-odd GDA---is suppressed relatively to the leading term by $\alpha_{\rm em}^2/\alpha_s^2$, yet features a distinct $\cos^2(\phi_P + \phi_K)$ dependence. The distinct azimuthal modulations of the various contributions provide a clean experimental handle for disentangling them. In particular, the moment $\langle\cos(\phi_{P}+\phi_{K})\rangle$ defined as
\begin{align}
   \int \! d\phi_P\, d\phi_K \cos(\phi_{P}+\phi_{K}) 
    \frac{{\rm d}\sigma^{+0;-0}}{{\rm d}s_{P}\,{\rm d}s_{K}\,{\rm d}u\,{\rm d}\zeta_{P}\,{\rm d}\phi_{P}\,{\rm d}\zeta_{K}\,{\rm d}\phi_{K}} \,,
    \label{e:cosphi_weighted}
\end{align}
isolates the interference term of chiral-even quantities ($\Psi_1$ and $\Psi_{2ce}$) with the chiral-odd ones $\Psi_{2co}$. This interference term depends on the phase difference between chiral-odd and chiral-even GDAs which enters as $\cos  (\delta_V^T(s_P)+\delta_V^T(s_K) - \delta_V(s_P)-\delta_V(s_K))$  . Since this phase difference vanishes at threshold due to Watson's theorem, we expect it to stay quite small in most of the $s_P,s_K$ range. This observable thus  enables a direct exploration of the chiral-odd GDA’s dependence on $(\zeta, s)$, once the chiral-even isovector GDA has been extracted from complementary processes such as deep electroproduction of a meson pair, $ep \to e' \pi^0 \pi^+ n$~\cite{Warkentin:2007su}.

\begin{figure}
\centering
\includegraphics[width=0.4\textwidth]{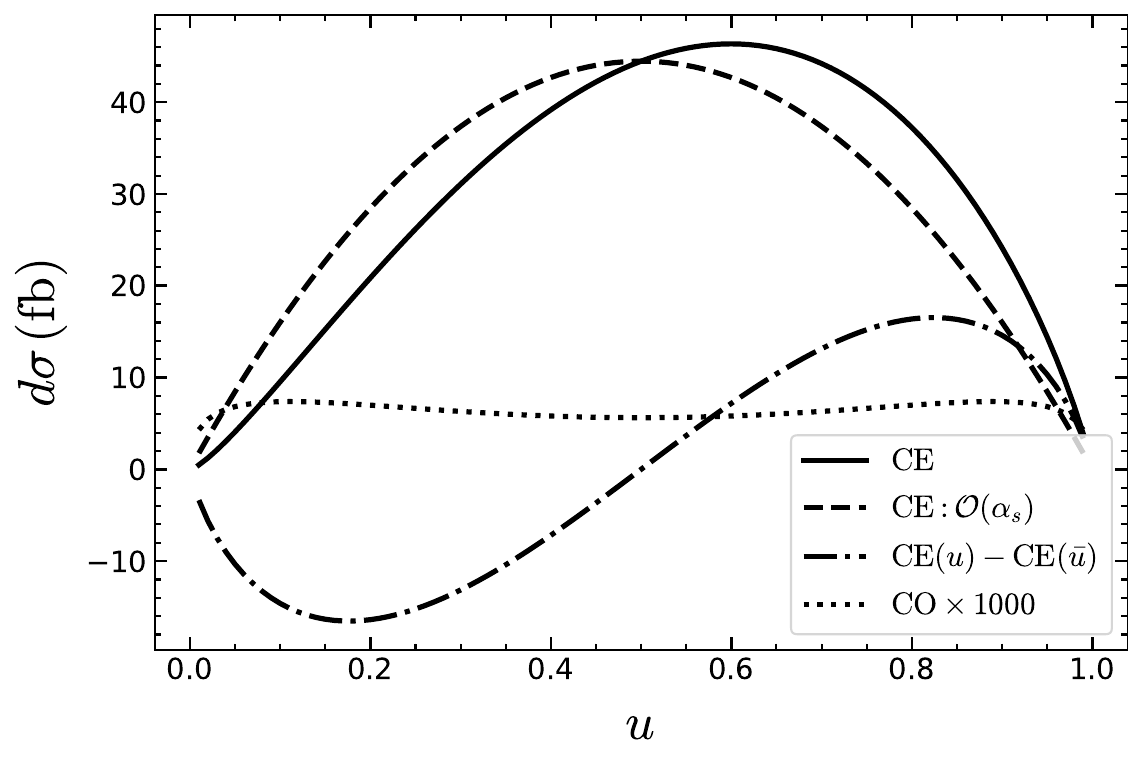}
\caption{
The pure chiral-even (CE) part of the cross section (solid line) at $s = 5~\text{GeV}^2$, integrated over $ \phi_P$ and $ \phi_K$, over $\zeta_P$, $\zeta_K$ in their full ranges, and over $s_P$ and $s_K$ between $4m_\pi^2$ and $1~\text{GeV}^2$. The leading single-photon exchange contribution (CE: $\mathcal{O}(\alpha_s)$) is shown by the dashed line. The dash-dotted curve represents the $u \to \bar{u}$ asymmetry arising from the interference between one- and two-photon exchange mechanisms within the chiral-even sector, i.e., ${\rm CE}(u) - {\rm CE}(\bar{u})$, and provides an estimate of the size of such effects. The (small) purely chiral-odd (CO) contribution, magnified by a factor of 1000 (CO $\times 1000$) for visibility, is shown by the dotted line.}
\label{fig:cross-section}
\end{figure}

\begin{figure}
\centering
\includegraphics[width=0.45\textwidth]{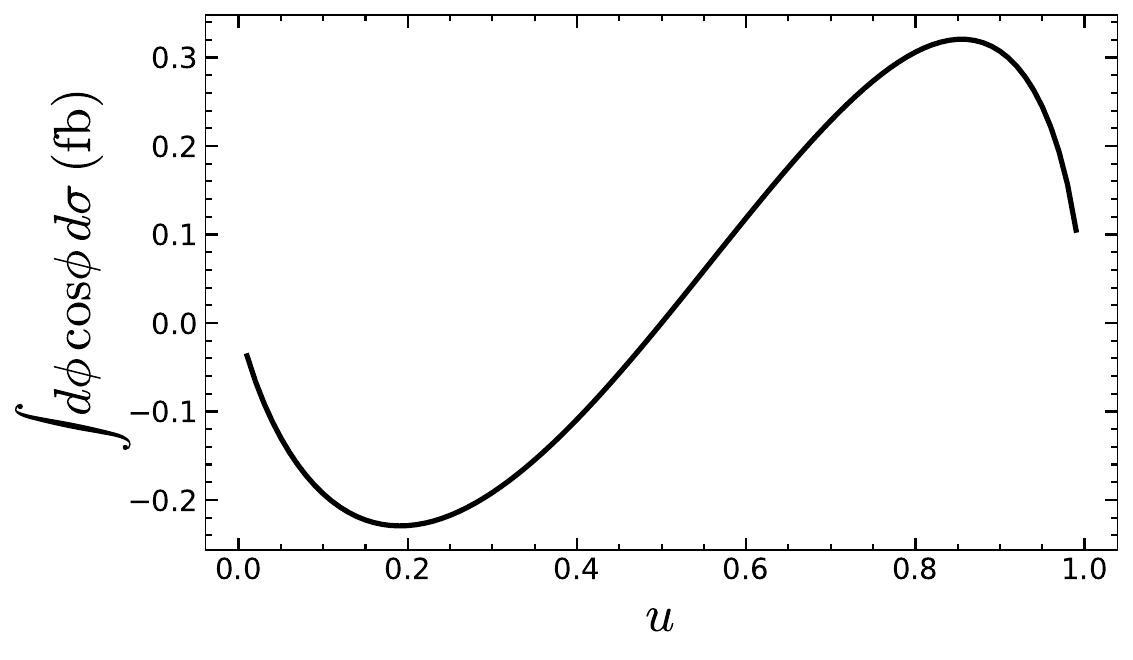}
\caption{The $\cos\phi$-weighted cross section, where $\phi \equiv \phi_P + \phi_K$, integrated over all variables except $u$, isolates the interference between chiral-odd and chiral-even contributions. The kinematics are chosen for $s = 5~\text{GeV}^2$.}
\label{fig:cosphiplot}
\end{figure}

To assess the experimental feasibility of the proposed measurement, we employ a simple model for the unknown chiral-odd GDAs, guided by existing knowledge of the chiral-even sector. Motivated by lattice results for the pion tensor form factor~\cite{Alexandrou:2021ztx}, we estimate the magnitude of the chiral-odd GDA using $B_{T10}(s, \mu) = 0.4\, B_{10}(s, \mu)$ as a benchmark.
Fig.~\ref{fig:cross-section} presents the various contributions to the differential cross section ${\rm d}\sigma/{\rm d}u$, integrated over the azimuthal angles $\phi_P$ and $ \phi_K$, the full ranges of $\zeta_P \in \left[\frac{1}{2} \left(1-\sqrt{1-\frac{4m_\pi^2}{s_P}}\right), \frac{1}{2} \left(1+\sqrt{1-\frac{4m_\pi^2}{s_P}}\right)\right]$,  
(idem for $\zeta_K$ with ($s_P\to s_K$))
and over $s_P$, $s_K \in [4m_\pi^2, 1~\text{GeV}^2]$. Note that apart from the limiting values of $\zeta_P,\zeta_K$, $s_P,s_K, \zeta_P,\zeta_K$ are independent variables. The upper limit of the $s_P,s_K$ integration range is arbitrary as long as it satisfies $s \gg s_K,s_P$ to justify the factorization of GDAs. These azimuthal angular integrations  remove the interference term in the second line of Eq.~(\ref{eq:dSigma+0-0}), arising from the chiral-even–chiral-odd interplay. As such, Fig.~\ref{fig:cross-section} displays only the first and last terms of Eq.~(\ref{eq:dSigma+0-0}). To isolate the suppressed but phenomenologically important interference term, we consider the $\cos(\phi_P + \phi_K)$-weighted cross section defined in Eq.~(\ref{e:cosphi_weighted}), whose result is shown in Fig.~\ref{fig:cosphiplot}. Despite its small magnitude, the signal's clean angular modulation may turn out to be accessible at high-luminosity $e^+e^-$ colliders.
Apart from running coupling constant and factorization scale dependence effects, all the contributions to the cross-section scale like $1/s^3$, the process will be more easily measurable at lower values of the collider energy. However, as can be seen from Eq.~(\ref{eq:dSigma+0-0}), the relative strength of the chiral-odd contributions compared with the chiral-even ones is essentially independent of 
$s$.
Once this chiral-odd chiral-even interference is isolated through the integrated $\cos(\phi_P + \phi_K)$-weighted cross section, a more precise extraction of the $(\zeta, s)$ dependence of chiral-odd GDAs becomes feasible, analogous to the analysis performed for chiral-even GDAs using Belle data~\cite{Kumano:2017lhr,Belle:2015oin}. Moreover, the relative phase between chiral-even and chiral-odd GDAs—neglected in our simplified parametrization—offers direct access to nonperturbative $\pi\pi$ final-state interactions, serving as a key probe of meson–meson dynamics and a benchmark for testing GDA models against phase-shift data.

 We note that observables odd under charge conjugation—such as forward-backward asymmetries, which have been employed in various contexts~\cite{Berger:2001xd,Hagler:2002nf,Hagler:2002nh,Pire:2008xe}—can also be used to isolate the interference between one-photon and two-photon  contributions.

{\bf 6. Prospects for a feasibility study at a future Super Tau–Charm Facility (STCF).} Let us  comment on the expected statistical reach at a future Super Tau-Charm Facility (STCF), following the design studies reported in Refs.~\cite{Achasov:2023gey, Ai:2025xop}. The STCF is envisioned to deliver an integrated luminosity of approximately $1~\mathrm{ab}^{-1}$ per year at $s = 13.5~\mathrm{GeV}^2$. As noted earlier, a somewhat lower center-of-mass energy would be optimal for the process under consideration; nevertheless, meaningful sensitivity can already be achieved at the nominal design point.

Focusing on the $(\pi^+\pi^0)(\pi^-\pi^0)$ channel and restricting to the nominal kinematic region at $s = 13.5~\mathrm{GeV}^2$, we estimate the event yield by integrating the cross section corresponding to the solid curve in Fig.~\ref{fig:cross-section}. This yields a total cross section of
\(\sigma = 1.53~\mathrm{fb}\), corresponding to approximately \(1.5\times 10^3\) events per year. While this represents a modest fraction of the total event sample, it is nevertheless statistically significant. With a dedicated analysis, these events can be isolated from dominant backgrounds, such as \(\tau^+\tau^-\) production followed by the dominant $\pi \pi^0 \nu$ decays, which are characterized by substantial energy loss due to neutrino emission.

Further insight is provided by the integrated charge or forward-backward $(u \leftrightarrow 1-u)$ asymmetry, evaluated over the interval $u\in[0.5,1]$. We find an integrated contribution of approximately $277.16~\mathrm{ab}$, corresponding to about 277 events per year. This clearly indicates that the interference between one-photon and two-photon exchange mechanisms should be experimentally observable.

To assess the sensitivity to chiral-odd GDAs, we translate the \(\cos\phi\) moment shown in Fig.~\ref{fig:cosphiplot} into an azimuthal asymmetry,
\begin{align}
\Delta \sigma
&=
\int_{0.5}^{1} du \int_{0}^{2\pi} d\phi_P
\int_{0}^{2\pi} d\phi_K
\Big[
\theta\big(\cos(\phi_P+\phi_K)\big)
\nonumber \\
& -
\theta\big(-\cos(\phi_P+\phi_K)\big)
\Big]
\nonumber\\
&\hspace{0.4cm}\times
\left[
\frac{d\sigma}{du\, d\phi_P\, d\phi_K}
-
\left.\frac{d\sigma}{du\, d\phi_P\, d\phi_K}\right|_{u\to 1-u}
\right] .
\end{align}
This yields an asymmetry of order $11.67~\mathrm{ab}$, corresponding to about 12 events per year. Compared to the 277 events discussed above, this is a $4$ per cent effect which is within reach of a careful data analysis. Note that this estimate is based on a conservative, non-maximal normalization of the chiral-odd GDAs. Larger normalizations would directly amplify this signal.
Finally, we emphasize that extending the data-taking period beyond a single year would substantially improve the statistical precision and further strengthen the feasibility of these measurements.

{\bf 7. Conclusions.} In this work, we proposed a novel strategy to access chiral-odd dimeson generalized distribution amplitudes (CO-GDAs), which parameterize the transverse-spin structure of mesons. GDAs provide the only portal to the three-dimensional tomography of mesons in the absence of physical meson targets, making them an indispensable counterpart to generalized parton distributions (GPDs) in mapping hadron structure. The chiral-odd sector, in particular, offers complementary access to spin-flip dynamics, including sensitivity to the pion’s anomalous magnetic moment through tensor current couplings absent in chiral-even observables.

We have demonstrated that CO-GDAs can be accessed in medium-energy $e^- e^+$ annihilation into two meson pairs, each with a relatively low invariant mass. Although the cross sections scale as $s^{-3}$ and are small, medium-energy, high-luminosity $e^+e^-$ colliders such as BES III or a future Super tau-charm facility (STCF) offer ideal conditions for their extraction~\cite{Achasov:2023gey}. Existing experimental datasets from BES III may already contain signals  for the exclusive four pion final state in the desired kinematics. 
We believe that a dedicated analysis could already measure the contribution coming from the interference of the one-photon and two-photon exchange processes. To be sensitive to the chiral-odd GDAs  will be an experimental challenge which we believe is worth the effort to uncover this elusive aspect of meson structure. Such a performance may be reached if  integrated luminosities for such a study reach the level of  1000 inverse femtobarns as expected for the future STCF. A more detailed feasibility study taking into account hadron identification efficiencies and a careful study of energy balance to reduce the background $\tau^+ \tau^-$ channel, obviously needs to be worked on, but this is beyond our scope here.
On the theoretical side, the chiral-odd and chiral-even GDAs merit investigation via lattice QCD, especially given recent advances in GPD studies; see for example~\cite{Bhattacharya:2025yba}. Nonperturbative models based on light-front wave functions~\cite{Dorokhov:2011ew,Wang:2022mrh} could also constrain these distributions, particularly in the soft-pion limit where the chiral-odd GDA connects to the twist-3 pion DA~\cite{Ball:1998je}. The present framework naturally extends to other channels, including different meson pairs or baryon--antibaryon final states~\cite{Han:2025mvq}.

We close by emphasizing that GDAs and GPDs represent two sides of the same nonperturbative coin, sharing a common spectral representation through double distributions~\cite{Radyushkin:1998es}. Yet, while GPDs have received considerable attention, GDAs remain largely unexplored. Since direct pion targets are experimentally inaccessible, exclusive two-meson production provides the only viable path to accessing both chiral-even and chiral-odd GDAs—offering essential input toward a quark and gluon understanding of meson structure, as emphasized throughout this work.


\begin{acknowledgements}
We acknowledge useful discussions with C\'edric Lorc\'e and C\'edric Mezrag. R.~B. was supported by the framework of the Saturated Glue (SURGE) Topical Theory Collaboration. L. S. was supported by the
Grant No. 2024/53/B/ST2/00968 and by the Grant No. 2019/33/B/ST2/02588 of  the National Science Centre in Poland.
\end{acknowledgements}
\bibliography{ref}

@article{Ai:2025xop,
    author = "Ai, Xiao-Cong and others",
    title = "{Conceptual design report of the Super Tau-Charm Facility: the accelerator}",
    eprint = "2509.11522",
    archivePrefix = "arXiv",
    primaryClass = "physics.acc-ph",
    doi = "10.1007/s41365-025-01833-x",
    journal = "Nucl. Sci. Tech.",
    volume = "36",
    number = "12",
    pages = "242",
    year = "2025"
}

@article{Bhattacharya:2025yba,
    author = "Bhattacharya, Shohini and Cichy, Krzysztof and Constantinou, Martha and Metz, Andreas and Miller, Joshua and Petreczky, Peter and Steffens, Fernanda",
    title = "{Generalized Parton Distributions from Lattice QCD with Asymmetric Momentum Transfer: Tensor Case}",
    eprint = "2505.11288",
    archivePrefix = "arXiv",
    primaryClass = "hep-lat",
    month = "5",
    year = "2025"
}

@article{Cheng:2019hpq,
    author = "Cheng, Shan",
    title = "{Dipion light-cone distribution amplitudes and $B \to \pi\pi$ form factors}",
    eprint = "1901.06071",
    archivePrefix = "arXiv",
    primaryClass = "hep-ph",
    doi = "10.1103/PhysRevD.99.053005",
    journal = "Phys. Rev. D",
    volume = "99",
    number = "5",
    pages = "053005",
    year = "2019"
}

@article{Belitsky:2005qn,
    author = "Belitsky, A. V. and Radyushkin, A. V.",
    title = "{Unraveling hadron structure with generalized parton distributions}",
    eprint = "hep-ph/0504030",
    archivePrefix = "arXiv",
    reportNumber = "JLAB-THY-04-34",
    doi = "10.1016/j.physrep.2005.06.002",
    journal = "Phys. Rept.",
    volume = "418",
    pages = "1--387",
    year = "2005"
}

@article{Diehl:2001pm,
    author = "Diehl, M.",
    title = "{Generalized parton distributions with helicity flip}",
    eprint = "hep-ph/0101335",
    archivePrefix = "arXiv",
    reportNumber = "DESY-01-009",
    doi = "10.1007/s100520100635",
    journal = "Eur. Phys. J. C",
    volume = "19",
    pages = "485--492",
    year = "2001"
}

@article{Lorce:2025ayr,
    author = "Lorc{\'e}, C{\'e}dric and Song, Qin-Tao",
    title = "{Spin-orbit correlation and spatial distributions for spin-0 hadrons}",
    eprint = "2501.05092",
    archivePrefix = "arXiv",
    primaryClass = "hep-ph",
    doi = "10.1016/j.physletb.2025.139433",
    journal = "Phys. Lett. B",
    volume = "864",
    pages = "139433",
    year = "2025"
}

@article{Brodsky:2002pr,
    author = "Brodsky, Stanley J. and Hwang, Dae Sung and Schmidt, Ivan",
    title = "{Single hadronic spin asymmetries in weak interaction processes}",
    eprint = "hep-ph/0211212",
    archivePrefix = "arXiv",
    reportNumber = "SLAC-PUB-9563, USM-TH-133",
    doi = "10.1016/S0370-2693(02)03259-8",
    journal = "Phys. Lett. B",
    volume = "553",
    pages = "223--228",
    year = "2003"
}

@article{Hagler:2009ni,
    author = "Hagler, Ph.",
    title = "{Hadron structure from lattice quantum chromodynamics}",
    eprint = "0912.5483",
    archivePrefix = "arXiv",
    primaryClass = "hep-lat",
    reportNumber = "TUM-T39-09-12",
    doi = "10.1016/j.physrep.2009.12.008",
    journal = "Phys. Rept.",
    volume = "490",
    pages = "49--175",
    year = "2010"
}

@article{Wang:2022mrh,
    author = "Wang, Xiaobin and Xing, Zanbin and Kang, Jiayin and Raya, Kh\'epani and Chang, Lei",
    title = "{Pion scalar, vector, and tensor form factors from a contact interaction}",
    eprint = "2207.04339",
    archivePrefix = "arXiv",
    primaryClass = "hep-ph",
    doi = "10.1103/PhysRevD.106.054016",
    journal = "Phys. Rev. D",
    volume = "106",
    number = "5",
    pages = "054016",
    year = "2022"
}

@article{Polyakov:1998ze,
    author = "Polyakov, Maxim V.",
    title = "{Hard exclusive electroproduction of two pions and their resonances}",
    eprint = "hep-ph/9809483",
    archivePrefix = "arXiv",
    reportNumber = "RUB-TPII-14-98",
    doi = "10.1016/S0550-3213(99)00314-4",
    journal = "Nucl. Phys. B",
    volume = "555",
    pages = "231",
    year = "1999"
}

@article{Dorokhov:2011ew,
    author = "Dorokhov, Alexander E. and Broniowski, Wojciech and Ruiz Arriola, Enrique",
    title = "{Generalized Quark Transversity Distribution of the Pion in Chiral Quark Models}",
    eprint = "1107.5631",
    archivePrefix = "arXiv",
    primaryClass = "hep-ph",
    doi = "10.1103/PhysRevD.84.074015",
    journal = "Phys. Rev. D",
    volume = "84",
    pages = "074015",
    year = "2011"
}

@article{Yan:2025ocu,
    author = "Yan, Da-Cheng and Li, Hsiang-nan and Rui, Zhou and Xiao, Zhen-Jun and Li, Ya",
    title = "{Improved global determination of two-meson distribution amplitudes from multi-body $B$ decays}",
    eprint = "2501.15150",
    archivePrefix = "arXiv",
    primaryClass = "hep-ph",
    month = "1",
    year = "2025"
}

@article{Ball:1998je,
    author = "Ball, Patricia",
    title = "{Theoretical update of pseudoscalar meson distribution amplitudes of higher twist: The Nonsinglet case}",
    eprint = "hep-ph/9812375",
    archivePrefix = "arXiv",
    reportNumber = "CERN-TH-98-400",
    doi = "10.1088/1126-6708/1999/01/010",
    journal = "JHEP",
    volume = "01",
    pages = "010",
    year = "1999"
}

@article{Alexandrou:2021ztx,
    author = "Alexandrou, Constantia and Bacchio, Simone and Cloet, Ian and Constantinou, Martha and Delmar, Joseph and Hadjiyiannakou, Kyriakos and Koutsou, Giannis and Lauer, Colin and Vaquero, Alejandro",
    collaboration = "ETM",
    title = "{Scalar, vector, and tensor form factors for the pion and kaon from lattice QCD}",
    eprint = "2111.08135",
    archivePrefix = "arXiv",
    primaryClass = "hep-lat",
    doi = "10.1103/PhysRevD.105.054502",
    journal = "Phys. Rev. D",
    volume = "105",
    number = "5",
    pages = "054502",
    year = "2022"
}

@article{Diehl:1998dk,
    author = "Diehl, M. and Gousset, T. and Pire, B. and Teryaev, O.",
    title = "{Probing partonic structure in gamma* gamma ---\ensuremath{>} pi pi near threshold}",
    eprint = "hep-ph/9805380",
    archivePrefix = "arXiv",
    reportNumber = "DAPNIA-SPHN-98-34, CPHT-S612-0598",
    doi = "10.1103/PhysRevLett.81.1782",
    journal = "Phys. Rev. Lett.",
    volume = "81",
    pages = "1782--1785",
    year = "1998"
}

@article{Diehl:2000uv,
    author = "Diehl, M. and Gousset, T. and Pire, B.",
    title = "{Exclusive production of pion pairs in gamma* gamma collisions at large Q**2}",
    eprint = "hep-ph/0003233",
    archivePrefix = "arXiv",
    reportNumber = "SLAC-PUB-8406, CPHT-S001-0100",
    doi = "10.1103/PhysRevD.62.073014",
    journal = "Phys. Rev. D",
    volume = "62",
    pages = "073014",
    year = "2000"
}

@article{Pire:2002ut,
    author = "Pire, B. and Szymanowski, L.",
    title = "{Impact representation of generalized distribution amplitudes}",
    eprint = "hep-ph/0212296",
    archivePrefix = "arXiv",
    doi = "10.1016/S0370-2693(03)00134-5",
    journal = "Phys. Lett. B",
    volume = "556",
    pages = "129--134",
    year = "2003"
}

@article{Belle:2015oin,
    author = "Masuda, M. and others",
    collaboration = "Belle",
    title = "{Study of $\pi^0$ pair production in single-tag two-photon collisions}",
    eprint = "1508.06757",
    archivePrefix = "arXiv",
    primaryClass = "hep-ex",
    reportNumber = "BELLE-PREPRINT-2015-15, KEK-PREPRINT-2015-24",
    doi = "10.1103/PhysRevD.93.032003",
    journal = "Phys. Rev. D",
    volume = "93",
    number = "3",
    pages = "032003",
    year = "2016"
}

@article{Hagler:2002nh,
    author = {H\"agler, Ph. and Pire, B. and Szymanowski, L. and Teryaev, O. V.},
    title = "{Hunting the QCD-Odderon in hard diffractive electroproduction of two pions}",
    eprint = "hep-ph/0202231",
    archivePrefix = "arXiv",
    reportNumber = "CPHT-S-001-01-02",
    doi = "10.1016/S0370-2693(02)01736-7",
    journal = "Phys. Lett. B",
    volume = "535",
    pages = "117--126",
    year = "2002",
    note = "[Erratum: Phys.Lett.B 540, 324--325 (2002)]"
}

@article{Hagler:2002nf,
    author = "Hagler, P. and Pire, B. and Szymanowski, L. and Teryaev, O. V.",
    title = "{Pomeron - odderon interference effects in electroproduction of two pions}",
    eprint = "hep-ph/0207224",
    archivePrefix = "arXiv",
    reportNumber = "CPHT-RR-060-0602",
    doi = "10.1140/epjc/s2002-01054-9",
    journal = "Eur. Phys. J. C",
    volume = "26",
    pages = "261--270",
    year = "2002"
}

@article{Warkentin:2007su,
    author = "Warkentin, N. and Diehl, M. and Ivanov, D. Yu. and Schafer, A.",
    title = "{Exclusive electroproduction of pion pairs}",
    eprint = "hep-ph/0703148",
    archivePrefix = "arXiv",
    reportNumber = "DESY-07-032",
    doi = "10.1140/epja/i2007-10374-9",
    journal = "Eur. Phys. J. A",
    volume = "32",
    pages = "273--291",
    year = "2007"
}

@article{Efremov:1979qk,
    author = "Efremov, A. V. and Radyushkin, A. V.",
    title = "{Factorization and Asymptotical Behavior of Pion Form-Factor in QCD}",
    reportNumber = "JINR-P2-12900",
    doi = "10.1016/0370-2693(80)90869-2",
    journal = "Phys. Lett. B",
    volume = "94",
    pages = "245--250",
    year = "1980"
}

@article{Lepage:1979zb,
    author = "Lepage, G. Peter and Brodsky, Stanley J.",
    title = "{Exclusive Processes in Quantum Chromodynamics: Evolution Equations for Hadronic Wave Functions and the Form-Factors of Mesons}",
    reportNumber = "SLAC-PUB-2343",
    doi = "10.1016/0370-2693(79)90554-9",
    journal = "Phys. Lett. B",
    volume = "87",
    pages = "359--365",
    year = "1979"
}

@article{Han:2025mvq,
    author = "Han, Jing and Pire, Bernard and Song, Qin-Tao",
    title = "{Baryon-antibaryon generalized distribution amplitudes and $e^+ e^- \to B \bar{B} \gamma$}",
    eprint = "2506.09854",
    archivePrefix = "arXiv",
    primaryClass = "hep-ph",
    month = "6",
    year = "2025"
}

@article{Berger:2001xd,
    author = "Berger, Edgar R. and Diehl, M. and Pire, B.",
    title = "{Time - like Compton scattering: Exclusive photoproduction of lepton pairs}",
    eprint = "hep-ph/0110062",
    archivePrefix = "arXiv",
    reportNumber = "CPHT-S010-0201, DESY-01-119",
    doi = "10.1007/s100520200917",
    journal = "Eur. Phys. J. C",
    volume = "23",
    pages = "675--689",
    year = "2002"
}

@article{Achasov:2023gey,
    author = "Achasov, M. and others",
    title = "{STCF conceptual design report (Volume 1): Physics {\&} detector}",
    eprint = "2303.15790",
    archivePrefix = "arXiv",
    primaryClass = "hep-ex",
    doi = "10.1007/s11467-023-1333-z",
    journal = "Front. Phys. (Beijing)",
    volume = "19",
    number = "1",
    pages = "14701",
    year = "2024"
}

@article{Omnes:1958hv,
    author = "Omnes, R.",
    title = "{On the Solution of certain singular integral equations of quantum field theory}",
    doi = "10.1007/BF02747746",
    journal = "Nuovo Cim.",
    volume = "8",
    pages = "316--326",
    year = "1958"
}

@article{Pire:2008xe,
    author = "Pire, B. and Schwennsen, F. and Szymanowski, L. and Wallon, S.",
    title = "{Hard Pomeron-Odderon interference effects in the production of $\pi^{+} \pi^{-}$ pairs in high energy gamma-gamma collisions at the LHC}",
    eprint = "0810.3817",
    archivePrefix = "arXiv",
    primaryClass = "hep-ph",
    reportNumber = "CPHT-RR078.1008, LPT-ORSAY-08-84",
    doi = "10.1103/PhysRevD.78.094009",
    journal = "Phys. Rev. D",
    volume = "78",
    pages = "094009",
    year = "2008"
}

@article{Kumano:2017lhr,
    author = "Kumano, S. and Song, Qin-Tao and Teryaev, O. V.",
    title = "{Hadron tomography by generalized distribution amplitudes in pion-pair production process $\gamma^* \gamma \rightarrow \pi^0 \pi^0 $ and gravitational form factors for pion}",
    eprint = "1711.08088",
    archivePrefix = "arXiv",
    primaryClass = "hep-ph",
    reportNumber = "J-PARC-TH-0086, KEK-TH-1959, J-PARC-TH-0086 (Erratum: KEK-TH-2096, J-PARC-TH-0155)",
    doi = "10.1103/PhysRevD.97.014020",
    journal = "Phys. Rev. D",
    volume = "97",
    number = "1",
    pages = "014020",
    year = "2018"
}

@article{Radyushkin:1998es,
    author = "Radyushkin, A. V.",
    title = "{Double distributions and evolution equations}",
    eprint = "hep-ph/9805342",
    archivePrefix = "arXiv",
    reportNumber = "JLAB-THY-98-16",
    doi = "10.1103/PhysRevD.59.014030",
    journal = "Phys. Rev. D",
    volume = "59",
    pages = "014030",
    year = "1999"
}

@article{Diehl:2003ny,
    author = "Diehl, M.",
    title = "{Generalized parton distributions}",
    eprint = "hep-ph/0307382",
    archivePrefix = "arXiv",
    reportNumber = "DESY-THESIS-2003-018",
    doi = "10.1016/j.physrep.2003.08.002",
    journal = "Phys. Rept.",
    volume = "388",
    pages = "41--277",
    year = "2003"
}

\end{document}